\documentclass[twocolumn,aps,pre,floatfix,superscriptaddress]{revtex4-1}%

\usepackage{amsmath,amsfonts,amssymb}
\usepackage[utf8]{inputenc}
\usepackage{bm}
\usepackage{enumerate}
\usepackage{epsfig}
\usepackage{float}
\usepackage{graphicx}
\usepackage{hyperref}
\usepackage{color}

\graphicspath{{./Fig_with_labels//}}

\def\ps{p_s}

\begin{document}

\title{What does Big Data tell? Sampling the social network by communication channels}

\author{János Török}
\email{torok@phy.bme.hu}
\affiliation{Department of Theoretical Physics, Budapest University of Technology and Economics, Budapest H-1111, Hungary}
\affiliation{Center for Network Science, Central European University, Budapest H-1051, Hungary}
\author{Yohsuke Murase}
\email{yohsuke.murase@gmail.com}
\affiliation{RIKEN Advanced Institute for Computational Science, Kobe, Hyogo 650-0047, Japan}
\author{Hang-Hyun Jo}
\email{johanghyun@postech.ac.kr}
\affiliation{Department of Physics, Pohang University of Science and Technology, Pohang 37673, Republic of Korea}
\affiliation{Department of Computer Science, Aalto University School of Science, P.O. Box 15500, Espoo, Finland}
\author{János Kertész}
\affiliation{Center for Network Science, Central European University, Budapest H-1051, Hungary}
\affiliation{Department of Theoretical Physics, Budapest University of Technology and Economics, Budapest H-1111, Hungary}
\affiliation{Department of Computer Science, Aalto University School of Science, P.O. Box 15500, Espoo, Finland}
\author{Kimmo Kaski}
\affiliation{Department of Computer Science, Aalto University School of Science, P.O. Box 15500, Espoo, Finland}

\begin{abstract}
Big Data has become the primary source of understanding the structure
and dynamics of the society at large scale. The network of social
interactions can be considered as a multiplex, where each layer
corresponds to one communication channel and the aggregate of all
of them constitutes the entire social network. However, usually one has
information only about one of the channels or even a part of it, which
should be considered as a subset or sample of the whole. Here we
introduce a model based on a natural bilateral communication channel
selection mechanism, which for one channel leads to consistent changes
in the network properties. For example, while it is expected that the
degree distribution of the whole social network has a maximum at a
value larger than one, we get a monotonously decreasing distribution
as observed in empirical studies of single channel data.  We also find
that assortativity may occur or get strengthened due to the sampling
method. We analyze the far-reaching consequences of our findings.
\end{abstract}

\date{\textrm{\today}}

\maketitle

\section{Introduction}\label{sect:intro}

Over the past decades the information-communication technology (ICT) has changed in various ways how we communicate and interact with each
other. Yet at the same time it has revolutionized social sciences~\cite{Lazer2009Computational} by making available an
unprecedented amount of high-quality data of social interactions of huge number of people. Through the computational analysis and
subsequent modeling one could get insight into earlier inaccessible properties like the structure of the interaction network at the
societal level~\cite{Onnela2007Structure}, the inhomogeneous dynamics of communication~\cite{Karsai2011Small, Iribarren2009Impact,
Jo2012Circadian}, and the laws of collective attention~\cite{Wu2007Novelty} to name a few examples. In these cases
the data is usually in the form of communication records, e.g., mobile phone calls, text messages, and emails~\cite{Onnela2007Analysis,
Wu2010Evidence, Eckmann2004Entropy}, as well as social networking services (SNS), e.g., Facebook and Twitter~\cite{Ugander2011Anatomy, Kwak2010What}.
While in each case one has information about a particular kind of interaction, the general interest stems also from the
assumption that this type of research can provide insight into the structure and function of the society as a whole. 

It is now understood that a network of human social interactions
should be considered as a multiplex
network where the each edge is categorized by its type~\cite{Boccaletti2014Structure,Kivela2014Multilayer,Lee2015towards}
at least from two different points of view. Usually one assumes that
the links can be classified according to the nature of relationships
like kinship, friendship, workmate links, etc. Each of these defines a
network, which then serves as a layer of the whole multiplex network.
On the other hand the interaction can also be assorted according to
the channels used for communication like face-to-face, mobile phone,
social network services, etc. Then the layers of the multiplex
correspond to different communication channels. Data is usually
available only for one channel, meaning that from the whole multiplex
there is only one layer we can investigate at a time. Linking data
from diverse channels is in most cases impossible due to their
different origins and for privacy reasons~\footnote{Exceptions are the
Reality Mining type data~\cite{Eagle2009Inferring,
Jo2012Spatiotemporal, Stopczynski2014Measuring}, where multichannel
information is collected from a relatively small group of volunteers.
This is clearly a very interesting line of research but is not
directly suitable for large scale conclusions.}.

\begin{figure}
\begin{center}
 \includegraphics[width=\columnwidth]{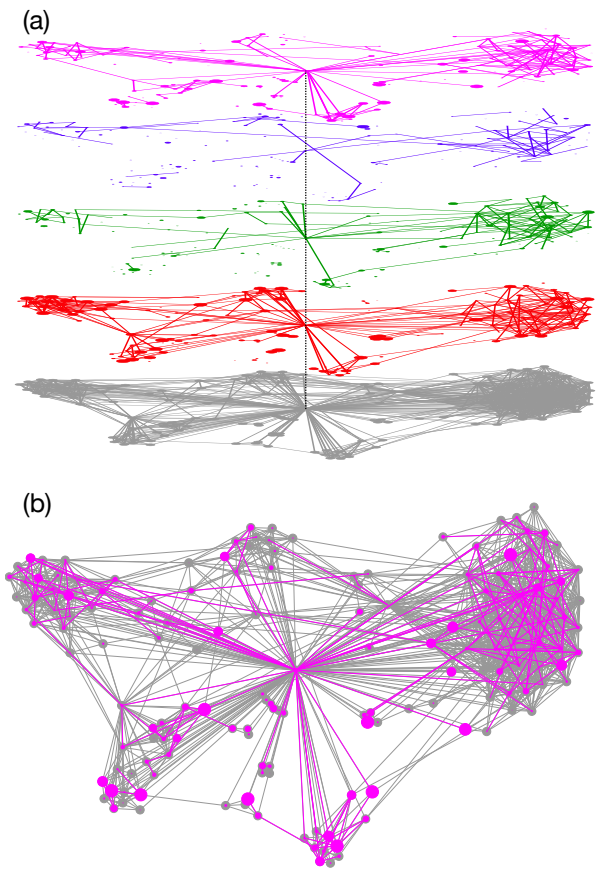}
 \caption{\label{Fig:sample}(Color online) (a) Schematic
 representation of the different communication channels or layers
 (different colors) on a sample egocentric network (gray). (b) One of
 the communication layers on the top of the original egocentric
 network. For the selection of the users, we used the sampling method
 defined in Section~\ref{sect:method} with $f_0=0.25$.}
 \end{center}
\end{figure}

Having information only about one layer of the multiplex raises the following questions. To what extent can we from the analysis of one particular layer draw  conclusions of the properties of the whole network? How much the properties of the whole network are reflected by the partial datasets? The answer to these questions is of fundamental importance, if we want to apply the results from the available data to the whole society. A schematic view of this picture is shown in Fig.~\ref{Fig:sample}. The most apparent problem here is the general observation that in ICT provided data about large populations, wherever they come from, the degree distribution shows a monotonously decreasing behavior~\cite{Onnela2007Structure,Ugander2011Anatomy,Newman2004Coauthorship}. This has the consequence that the most probable degree is one. Then a simple-minded generalization of this observation would imply that this statement is true for a number of social contacts of an individual, which is clearly nonsense. 

Dealing with data from only one communication channel can be
interpreted as sampling such that each layer constitutes a sample of
the whole multiplex network, comprised of the people using a
particular channel of communication. The sampling method changes the
properties of the network and it is an inverse
problem~\cite{Aster2012Parameter} to draw conclusions about the whole
system from the partial observations. 

We would like to stress the difference from previous studies on
sampling networks~\cite{Stumpf2005subnets, Stumpf2005sampling,
Lee2006Statistical}. These studies focused on the bias caused by
selecting a fraction of the data (or network) for analysis such that
its statistical properties remain unaltered. While this sampling is a
statistical issue and has previously been studied well, the selection
of social links by a single communication channel is inherently
social activity and has so far not attracted much research attention.

Generally speaking, available datasets usually undergo two-step
sampling \cite{gjoka2010walking}. The first sampling takes place when
the calling person chooses a communication channel from various
options depending on context or a person with whom the contact is
made. The other sampling occurs when an observer or a researcher
analyzes the dataset.  Since it is often hard to analyze all the logs
in an ICT service mainly due to technical reasons, a fraction of data
are randomly sampled for the analysis and the properties are
statistically inferred. In this paper, we will focus on the former
sampling, which is of fundamental importance since a bias originating
from social activity is not reduced by the amount of data.

In this paper we analyze the relationship between the whole
network and the sampled network using different techniques. We will show
how the sampling may substantially change the properties of the
network, e.g., it can make a monotonic degree distribution from one
with a peak at degree larger than one.  We will demonstrate under
which condition the sampled network reflects the properties of the
surrogate network.

\section{Empirical observations}\label{sect:empirical}

\begin{figure*}
\begin{center}
\includegraphics[width=0.85\textwidth]{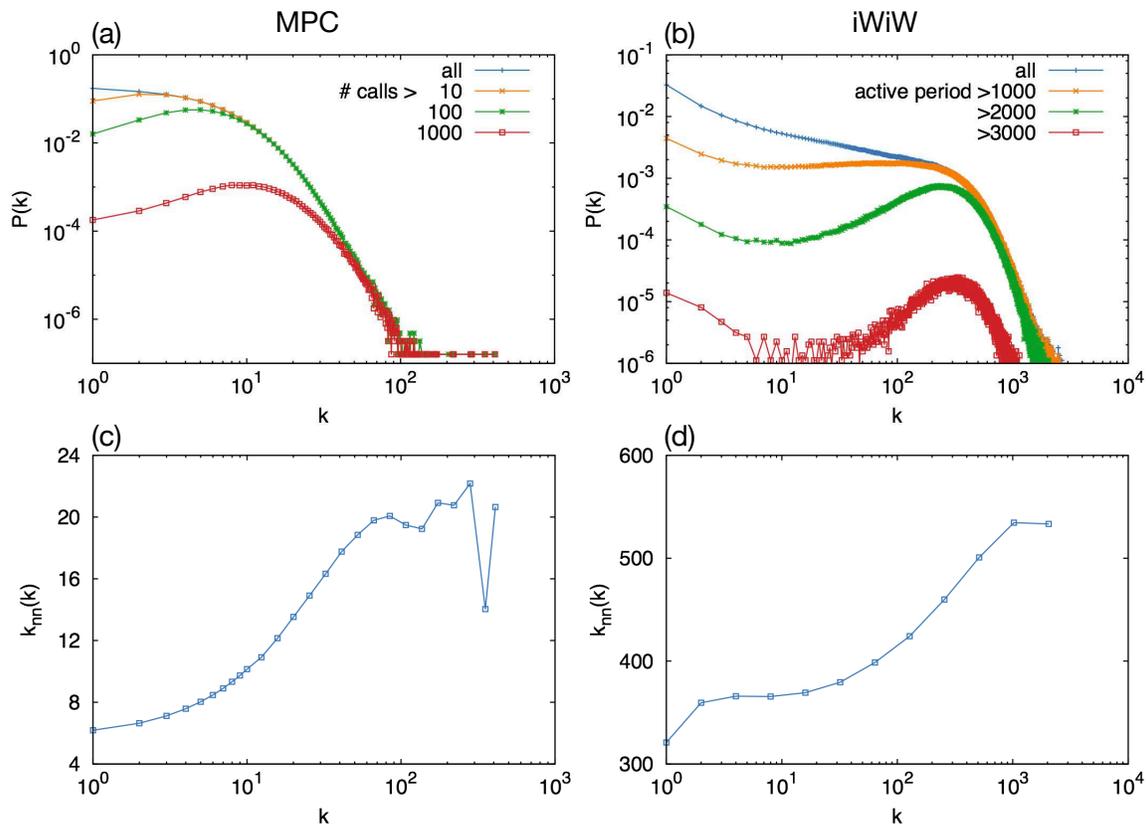}
\caption{\label{Fig:empirical}
  (Color online)
  Empirical results of the mobile phone call (MPC) dataset (left) and the Hungarian social networking service iWiW dataset (right).
  Degree distributions for users with activity larger than some threshold values (top) and the average degree of neighbors of users with degree $k$ (bottom).
  For the degree distributions, frequencies are divided by the total number of nodes in each dataset, i.e., the curves for ``all'' are normalized while the others are not.
  }
\end{center}
\end{figure*}

We first summarize some general empirical findings or \emph{stylized facts} in social networks from different ICT datasets~\cite{Onnela2007Analysis, Kwak2010What, Ugander2011Anatomy, Lengyel2015Geographies}. Here we have chosen to analyze two datasets: One is the mobile phone call (MPC) dataset~\cite{Onnela2007Structure} and the other is the Hungarian social networking service iWiW~\cite{Lengyel2015Geographies}. For the MPC network, we consider a link between two users existing provided that it is mutual~\cite{Onnela2007Structure}, and for the iWiW network when the friendship was recognized by both users.

Figure~\ref{Fig:empirical} summarizes degree distributions $P(k)$ and  the average degree of neighbors $k_{nn}(k)$ of the users with degree $k$, which are measured for the MPC and iWiW networks, respectively. For both networks, the degree distribution is found to decrease monotonically. This decrease is slow for a range of small $k$, then followed by a fast decay that is either exponential or follows a power-law with very large decay exponent~\cite{Onnela2007Structure}. Both networks show assortativity, characterized by an increasing $k_{nn}(k)$ as a function of $k$. In this paper we focus on these two properties.

The decreasing $P(k)$ and the increasing $k_{nn}(k)$ are generally observed for ICT datasets, despite their diverse origins. For example, the average degree for the MPC network is small, i.e., of the order of $10$, while for the Facebook network it is large, i.e., of the order of $100$~\cite{Ugander2011Anatomy}. Furthermore, the growth mechanisms of the networks differ from each other. Some require invitation, e.g., in case of iWiW, while others require paid subscription as in case of MPC. However, the stylized facts across various datasets imply that there could exist a common underlying mechanism.

It is reasonable to assume that very few people have only one social contact, thus the maximum of the distribution should not be at unity 
~\cite{hill2003social}. This implies that the degree distribution should first increase and then decrease, which we call a peaked degree distribution. We argue that the discrepancy between this plausible picture and most empirical findings from the ICT datasets can be attributed to the selection method of a single communication channel from the whole network of social interaction between people. 

In order to support our motivation, we briefly argue the following. As it takes time and effort for people to build up a network on a communication channel, people with larger activity may develop their egocentric networks in a particular channel more similarly to their real egocentric networks. For example, for the MPC network, an activity of a user can be defined as the total number of calls. Then the degree distribution only for users with activity larger than some threshold value is expected to be more similar to the real degree distribution. Figure~\ref{Fig:empirical}(a) shows that the degree distributions for users with activity above sufficiently large thresholds show peaked behavior, whereas the degree distribution for all users is monotonically decreasing. This may indicate that the discrepancy of overweighting the low degree nodes comes from the low activity users. Since we do not have activity records for the iWiW network, as a proxy for it we use the active period defined by the number of days between the first and last logins. We find the same transformation from monotonically decreasing to peaked degree distribution, as shown in Fig.~\ref{Fig:empirical}(b). As the egocentric networks of users with large activity are expected to be more similar to the real egocentric networks of users, the observed peaked distribution for such users can be considered as evidence for the peaked degree distributions in the entire social network.

Another important property of social networks is assortative mixing, which is usually attributed to the link formation mechanism related to homophily~\cite{McPherson2001Birds}. As we do not know the underlying social network exactly, we cannot confirm this at the societal scale. However, we will consider this property here. 

\section{Sampling Method}\label{sect:method}

It is important to stress that ICT datasets report dynamics of a single communication channel while people in general use many different means of communication. The natural relationship network of humans is thus a multi-layer system, in each layer of which there are only links that represent a single communication channel. Therefore, the data related to a single layer can be considered as a special sampling of the entire network. In order to understand this sampling we model how people choose a communication channel or an ICT service to make contacts with other people. 

People have diverse interests and preferences for communication and they show different usage patterns of services such as the frequency of visits or the variation of the time spent using the service. For example, users that have invested a considerable amount of time to build up their friendship network would show a higher preference to use the service. The degree of preference of a user to choose a service can be described by an \emph{affinity} quantity, denoted by $f_i$ for user $i$. 

Let us now assume that agents $i$ and $j$ know each other, and they try to communicate and for that purpose they have to choose a communication channel. In general the agents have different personal affinities towards different communication channels. When they are choosing a channel they tend to avoid those that are inconvenient for the other, because it could risk the success of communication. For example, writing an urgent email to someone who is checking it weekly is not a good idea. Alternatively waiting for someone to appear on an instant messaging channel, if login is irregular, could be meaningless. So naturally everyone tends to choose those channels to communicate with an acquaintance for which both of them have relatively high affinity. Hence, we assume the probability that a link between $i$ and $j$ is made over a given communication channel is a symmetric function of $f_i$ and $f_j$, $p_{ij}(f_i, f_j)$, as introduced in \cite{caldarelli2002scalefree,boguna2003class}.

Our strategy to investigate the effect of sampling is as follows.
Since the real, underlying social network is unknown, we generate {\em
surrogate networks} with the given properties, such as peaked degree
distributions. To these networks we apply a sampling method that
mimics the usage of a single communication channel. We note here
that ICT dataset may cover only part of the population due to
competing services. Random selection of nodes or links does not change
the basic characteristics of the degree distribution whether it is 
monotonously decreasing or peaked.

We assume that the affinity distribution $P(f)$ is a decreasing
function (e.g. \cite{Onnela2007Structure,gyarmati2010measuring}). This
implies that a large fraction of people rarely spend time using the
service, while there are a relatively small number of enthusiastic
users. As for the affinity distribution, we choose an exponential
function 
\begin{equation}\label{Eq:expdist}
    P(f)=\frac{1}{f_0}e^{-f/f_0},
\end{equation}
where $f_0$ is the average affinity and it serves as a control
parameter. Each user is assigned an affinity value that is randomly
drawn from $P(f)$, which implies that the correlation between
affinities of neighboring users is ignored for simplicity. The effect
of such correlations existing in reality can be studied for future
work.

We then sample links in the surrogate network with a probability as a function of affinities of users connected by the link. For neighboring users $i$ and $j$, the probability of sampling a link $ij$ is defined as follows
\begin{equation}\label{Eq:minrule}
    p_{ij}=\min\{f_i,f_j,1\}.
\end{equation}
The set of sampled users consists of users that have at least one sampled link. An example of a sampled network from an egocentric network is presented in Fig.~\ref{Fig:sample}.
We would like to note that our model takes into account only two
point correlations. As expected, then all higher order correlations
such as clustering are systematically lost by this sampling.

Here we assume the minimum rule for $p_{ij}$ since a link is often established in a communication channel when both nodes $i$ and $j$ accept to use it. As we will see in Section~\ref{sect:gen_mean}, this rule is not only a natural consequence of mutual acceptance, but it is the most representative rule for a broader class of rules that reproduce the observed stylized facts. In addition, the minimum rule is analytically solvable as shown in the next Section.

We now consider three kinds of surrogate networks: Random regular graphs (RR) with degree $k_0$, Erd\H{o}s-R\'enyi random graphs (ER) with average degree $\langle k\rangle$, and weighted social networks using link deletion (WSN) with average degree $\langle k\rangle$. We will use the WSN studied in our previous work~\footnote{The parameters to generate WSN are $N=10^4$, $p_{\Delta}=0.07$, $p_r=0.0007$, $p_{ld}=0.0015$, and the maximum time step $t_{max}=50000$. The notation of the parameters are the same as in~\cite{Murase2015Modeling}.}. All three networks show peaked degree distributions. The RR and ER do not show assortative mixing, while the WSN was devised to produce high clustering, community structure, and assortative mixing, as observed in real networks based on ICT data.

\section{Results}\label{sect:results}

\subsection{Degree distribution}

\begin{figure}
\begin{center}
\includegraphics[width=\columnwidth]{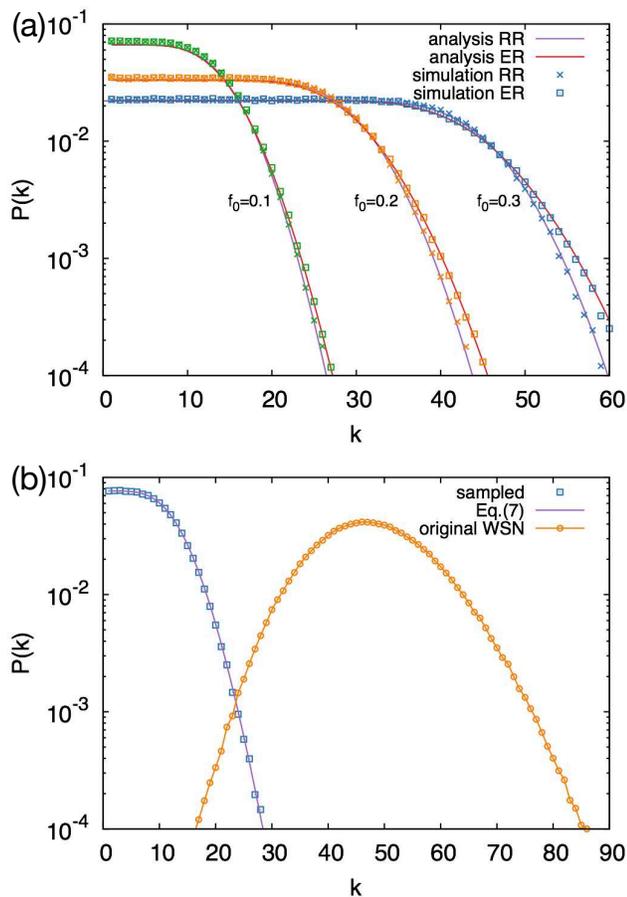}
\caption{\label{Fig:degreedist}
    (Color online) Degree distributions of the sampled networks with
    analytic curves obtained using Eq.~(\ref{eq:Pk}). (a) The
    surrogate networks are regular random graphs (RR, $\times$) or
    Erd\H{o}s-R\'{e}nyi random graphs (ER, $\Box$) with a network size
    of $N=10^4$ and the degree of $k_0=150$ (RR) or the average degree
    of $\langle k\rangle=150$ (ER). Analytic solutions for several
    values of $f_0$ (curves) perfectly fit the simulation results
    (symbols). The simulation results are averaged over $50$
    independent runs. (b) The surrogate network is a weighted social
    network using the link deletion model of
    Ref.~\cite{Murase2015Modeling} with parameters $N=10^4$ and
    $\langle k\rangle\approx 47.8$. Using $f_0=0.3$, we find that the
    size of the sampled network is $9287$ and that its average degree
    is $\langle k\rangle\approx 7.7$. Since the surrogate network has
    a high average degree, most of the nodes remain in the sampled
    network even after majority of links are removed. }
\end{center}
\end{figure}

Monotonically decreasing degree distributions are found for sampled networks in all the surrogate networks, as depicted in Fig.~\ref{Fig:degreedist}. For the RR case, degree distributions of the sampled networks are flat up to the crossover degree $\approx f_0k_0$, followed by exponentially decaying behavior. Similar patterns are observed for the ER and WSN cases with crossover degrees $\approx f_0\langle k\rangle$.

We can analytically calculate the degree distribution of the network sampled from the regular random graph with degree $k_0$. Since the affinities between neighboring nodes are uncorrelated, the probability of sampling a link involving the node $i$ with affinity $f_i$ is obtained as
\begin{equation}\label{Eq:r_fi}
    \ps(f_i)=\int_0^\infty
    p_{ij}P(f_j)df_j=f_0\left(1-e^{-\min\{f_i,1\}/f_0}\right).
\end{equation}
Then we obtain the probability that the node $i$ has exactly $k_i$ links in the sampled network as
\begin{equation}
q(k_i|k_0,f_i)=\binom{k_0}{k_i}\ps(f_i)^{k_i} [1-\ps(f_i)]^{k_0-k_i}.
\end{equation}
The degree distribution of the sampled network is calculated as
\begin{align}\label{Eq:pkanal}
    Q_{k_0}(k)&=\int_0^\infty q(k|k_0,f_i)P(f_i) df_i \cr
  &\approx \frac{1}{f_0(k_0+1)}I_{\left(\frac{f_0}{1-f_0}\right)}(k+1,k_0-k+1),
\end{align}
where $I_x(a,b)$ denotes the regularized beta function. Here we used the approximation that $\min\{f_i,1\}=f_i$ for all $i$, which is the case for the sufficiently small value of $f_0$. This analytical solution perfectly fits the simulation results, as shown in Fig.~\ref{Fig:degreedist}(a).

The first part of the degree distribution is flat, which can be calculated. The function $q(k_i|k_0,f_i)$ may have a very strong peak and can be approximated by a Dirac delta function:
\begin{equation}
q(k_i|k_0,f_i)\simeq\delta(k_0 \ps(f_i) -k_i)
\end{equation}
which gives rise to constant $P(k)$ up to $k_0f_0$, and after which it is zero. In Appendix~\ref{sect:other_Pf}, other $P(f)$ functions are analyzed to show that the exponential function is a borderline between the case when $Q_{k_0}(k)$ is always decreasing and the case when it has a peak at $k>1$.

We now consider the case of ER graphs. The degree distribution of the surrogate network is binomially distributed, denoted by $P_0(k)$. The probability that a node originally having $k_0$ links will keep $k$ links is $Q_{k_0}(k)$ independently of the rest of the network so we can get the degree distribution for any uncorrelated network by a weighted sum of Eq.~(\ref{Eq:pkanal}):
\begin{equation}
    \label{eq:Pk}
    P(k)=\sum_{k'=0}^\infty P_0(k')Q_{k'}(k).
\end{equation}
We calculate $P(k)$ numerically to compare it with the simulation results, as shown in Fig.~\ref{Fig:degreedist}(a). Similarly, one can obtain $P(k)$ of the sampled network in the case of WSN [see Fig.~\ref{Fig:degreedist}(b)].

So far we have considered surrogate networks with degree distributions  decaying faster than exponential, whereas heavy-tailed degree distributions are observed in many ICT datasets. In order to consider more realistic situations, we generate surrogate networks with log-normal and L\'evy distributions of degree, where we used $\mu=\ln 200$ and $\sigma= \ln 2$ for log-normal distribution, and $\mu=0$ and $c=150$ for L\'evy distribution. Both distributions have peaks at values larger than $1$. Then, using Eq.~(\ref{eq:Pk}), we find that degree distributions in the sampled networks using $f_0=0.1$ are heavy-tailed but yet monotonically decreasing.

Based on the above analysis and simulations we conclude that the monotonically decreasing degree distribution in most ICT datasets could be the consequence of the sampling method applied to the real social network showing the peaked degree distribution. Our conclusion is robust with respect to the variation of details of the method as will be shown in Sec.~\ref{sect:gen_mean}.

\begin{figure}
\begin{center}
\includegraphics[width=\columnwidth]{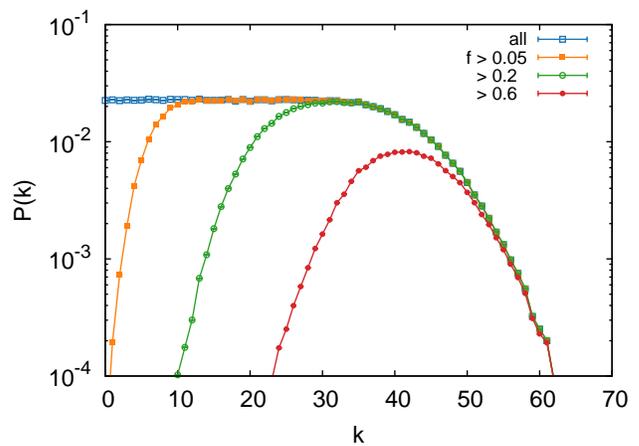}
\caption{\label{Fig:ER_pk_f} (Color online) The degree distributions in the sampled networks from ER for nodes with affinity larger than the indicated values.}
\end{center}
\end{figure}

We have found a mechanism transforming the ground-truth peaked degree
distribution to the observed monotonically decreasing one. This raises
the question whether the activity thresholding that resulted in an
opposite direction, i.e., from monotonically decreasing degree
distribution to the peaked one would work also for the model. For
this, we need to find an appropriate proxy for activity, and affinity
seems to be a good candidate since low (high) activity, i.e., low
(high) preference for a channel would imply low (high) activity on
that channel. We indeed find peaked degree distributions in the
sampled networks when considering only nodes with affinity larger than
some threshold value in Fig.~\ref{Fig:ER_pk_f}. This peaked behavior
compares favourably with the empirical observations as shown in
Fig.~\ref{Fig:empirical}, both in a linear or logarithmic scale. 

\begin{figure}
\begin{center}
\includegraphics[width=\columnwidth]{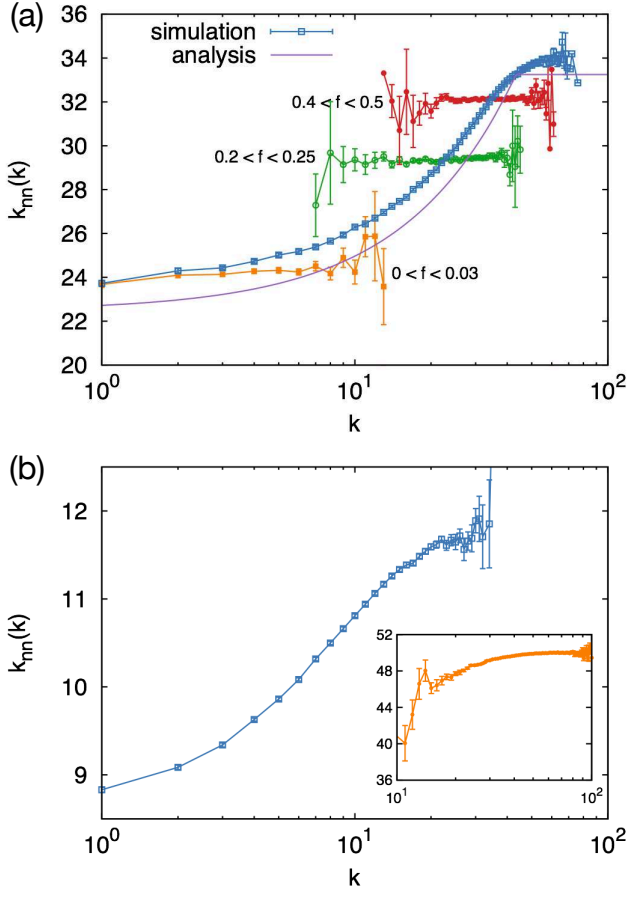}
\caption{\label{Fig:knn}
    (Color online) Average degrees of neighboring nodes $k_{nn}(k)$ as a function of the degree of the node in the sampled networks. (a) The surrogate networks are Erd\H{o}s-R\'{e}nyi random graphs used in Fig.~\ref{Fig:degreedist}(a). Blue squares show the result for all nodes, while others show the results only for nodes whose affinity is in the annotated ranges, respectively. Analytic solutions in Eqs.~(\ref{eq:knn_smallk}--\ref{eq:knn_largek}) are also depicted for comparison. (b) The assortativity $k_{nn}(k)$ for the network sampled from a weighted social network with link deletion used in Fig.~\ref{Fig:degreedist}(b). In the inset, $k_{nn}(k)$ for the surrogate network is plotted for comparison. In all cases, the simulation results are averaged over $50$ independent runs.
  }
\end{center}
\end{figure}

\subsection{Assortativity}\label{subsec:assortativity}

Here we investigate the effect of the proposed sampling method on the assortative mixing of the sampled networks. Figure~\ref{Fig:knn}(a) shows that the assortative mixing turns out to be present 
in the sampled networks even when the nodes in the surrogate network are completely uncorrelated. When the surrogate network shows assortative mixing, e.g., in the case of WSN, the assortativity is observed as expected, see Fig.~\ref{Fig:knn}(b).

We first calculate the correlation of affinities between neighboring nodes in the sampled networks. Similarly to the definition of $k_{nn}(k)$, we define the average affinity of neighboring nodes of a node $i$ with affinity $f_i$, denoted by $f_{nn}(f_i)$. Let us denote by $a_{ij}=1$ the event that a link $ij$ in the surrogate network is sampled. The affinity distribution of $f_j$ for the neighbor $j$ of the node $i$ in the sampled network can be written as a conditional probability $P(f_j|f_i,a_{ij}=1)$. Then we get
\begin{align}\label{Eq:fnn}
    f_{nn}(f_i)&=\int_0^\infty f_j 
    P(f_j|f_i,a_{ij}=1)
    df_j \cr
    &=\frac{1}{\ps(f_i)} \int_0^\infty f_j P(f_j) p_{ij} df_j \cr
    &=2f_0 - \frac{\min\{f_i,1\}}{e^{\min\{f_i,1\}/f_0}-1},
\end{align}
which turns out to be an increasing function of $f_i$. Note that $\ps(f_i)$ in Eq.~(\ref{Eq:r_fi}) is the probability of the event that $a_{ij}=1$ for a given affinity $f_i$. This result holds irrespective of a structure of the surrogate network. The positive correlation between affinities of neighboring nodes appears even when there is no such correlation in the surrogate networks.

This correlation is expected to persist also in $k_{nn}(k)$ since $k$ and $f$ are positively correlated. For the RR case with degree $k_0$, $k_{nn}$ for a node $i$ with affinity $f_i$ can be obtained as follows:
\begin{equation}\label{Eq:knn_general}
    k_{nn}(f {=} f_i) =\frac{k_0}{\ps(f_i)} \int_0^\infty \ps(f_j)p_{ij}P(f_j)df_j.
\end{equation}
The integral part is exactly solved using Eqs.~(\ref{Eq:expdist}--\ref{Eq:r_fi}) as
\begin{equation}
    f_0^2 \left(\frac{3}{4}-e^{-a_i/f_0}+ \frac{1}{4}e^{-2a_i/f_0} -\frac{a_i}{2f_0}e^{-2/f_0} \right),
\end{equation}
where $a_i\equiv \min\{f_i,1\}$. Since the expected degree of a node
with affinity $f$ is $k_0r(f)$, we replace $f$ in
Eq.~(\ref{Eq:knn_general}) using the assumption of $\ps(f)=\frac{k}{k_0}$ to get for $k<k_0f_0$:
\begin{equation}
    \label{eq:knn_smallk}
    k_{nn}(k)=\frac{k_0f_0}{2}+\frac{k}{4}+\frac{k_0^2f_0^2e^{-2/f_0}}{2k}\ln\left(1-\frac{k}{k_0f_0}\right),
\end{equation}
or otherwise 
\begin{equation}
    \label{eq:knn_largek}
    k_{nn}(k)=k_0f_0\frac{\frac{3}{4}- e^{-1/f_0} +\left(\frac{1}{4}-\frac{1}{2f_0}\right)e^{-2/f_0}} {1 - e^{-1/f_0}}.
\end{equation}
From this solution, one can obtain the extreme values of $k_{nn}(k)$, i.e., when the degree or the affinity is extremely small or large. If the affinity of a node $i$ is very small, we get $k_{nn}\approx
\frac{k_0f_0}{2}$. On the other hand, if the affinity is very large, one gets $k_{nn}\approx \frac{3k_0f_0}{4}$. We confirm numerically 
that these solutions apply also to the ER case, as shown in Fig.~\ref{Fig:knn}(a).

In Fig.~\ref{Fig:knn}(a), we have plotted $k_{nn}(k)$ for nodes whose $f$ is in a given range. The assortative behavior mostly disappears as in the surrogate network, implying that the assortativity in the sampled network is attributed to the dependence of $k_{nn}$ on $f$ but not on the assortativity of the surrogate network.

Assortative mixing is also observed for the networks sampled from the WSN, as shown in Fig.~\ref{Fig:knn}(b). This is not surprising because the surrogate network shows already assortativity. We note that $k_{nn}(k)$ for the surrogate network is concave, while the sampled network shows a slightly convex curve in the log-linear plot. We think that sampling enhances assortativity as compared to that of the surrogate network. This implies that the sampling plays a crucial role in the observed assortativity. An important lesson from this study is that the assortativity observed from sampling does not assure that the original (multiplex) network is assortative.

\subsection{Node strength}\label{subsec:othermeasures}

\begin{figure}
\begin{center}
\includegraphics[width=\columnwidth]{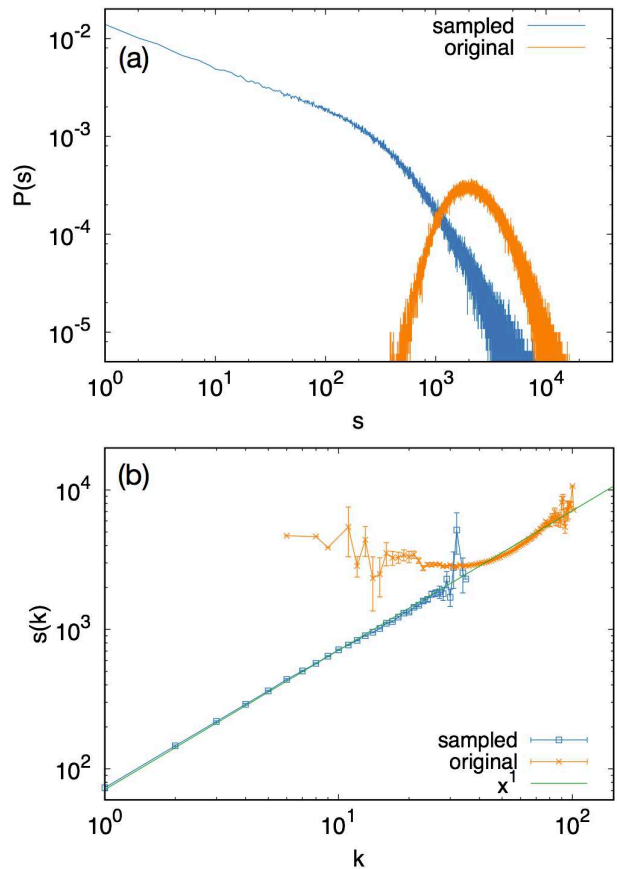}
\caption{\label{Fig:other} (Color online) The effects of sampling on
(a) the node strength distribution and (b) the average node strength
as a function of the degree when the WSN is used for the surrogate
network. Parameter values for the WSN are the same as in
Fig.~\ref{Fig:degreedist}(b).}
\end{center}
\end{figure}

The effect of sampling on the strength distribution of nodes and the
strength-degree correlation was tested using the WSN as a surrogate
network. The original strength distribution is peaked as seen in
Fig.~\ref{Fig:other}(a), and the strength-degree correlation in
Fig.~\ref{Fig:other}(b) shows a rather flat and then increasing behavior
which is the result of the WSN model \cite{Murase2015Modeling}.

After the sampling the low degree nodes in the sampled network are not
necessarily the ones that had low degree in the original network. The
value of affinity has much larger influence on the sampled degree of a
node than its original degree. Thus the link weight will be largely
independent of the sampled degree of the nodes the link is connected
to. This implies linear relationship between the degree and strength, which
is exactly what was found in Fig.~\ref{Fig:other}(b). Similar
relation was found in the mobile phone dataset presented in
\cite{Onnela2007Analysis}, where a correlation $s\propto k^\gamma$
with $\gamma\simeq 0.8{-}0.9$ was obtained. The empirical data shows
almost linear strength-degree correlation, but our model indicates that
this result has no implications for the original social network.

As there is only marginal correlation between the degree and the link
weight, the degree distribution mostly determines the strength
distribution, which is the convolution of the link weight and degree
distributions.  As a result the strength distribution is a decreasing
function instead of being peaked as shown in Fig.~\ref{Fig:other}(a).

\section{Generalization of the Model}\label{sect:gen_mean}

In this section, we generalize our model and discuss the robustness of the results we have seen in the previous section. We will focus on the effects of the affinity distribution and sampling probability.

\subsection{Generalized-Mean Model}\label{subsect:gen_mean_model}

First, we generalize the distribution of affinity, $P(f)$, to be a truncated Weibull distribution:
\begin{equation}\label{eq:pf_general}
  P(f) = \begin{cases}
      c(f/f_0)^{\alpha-1}e^{-(f/f_0)^\alpha} & \mathrm{if~} 0 \leq f\leq 1  \cr
      0 & \mathrm{otherwise}
    \end{cases}
\end{equation}
where $c$ is the normalization constant. Here we implement the truncation at $f=1$ in the affinity distribution instead of the sampling probability. Obviously $\alpha=1$ gives back the exponential distribution.
When $\alpha<1$, a divergence at $f=0$ is seen thus the nodes tend to have a smaller $f$.

The sampling probability $p_{ij}$ is defined as a generalized mean of $f_i$ and $f_j$ with exponent $\beta$:
\begin{equation}\label{eq:pij_general}
  p_{ij} = \begin{cases}
      \left( \frac{f_i^\beta + f_j^\beta}{2} \right)^{1/\beta} &
      \mathrm{if~} \beta \neq 0 \cr
      \sqrt{f_if_j} & \mathrm{if~} \beta = 0.
    \end{cases}
\end{equation}
The generalized mean includes Pythagoranean means as special cases such that for example when $\beta$ is $1$, $0$, or $-1$, it is equivalent to arithmetic, geometric, or harmonic mean, respectively.  For any real $\beta$, $p_{ij}$ is an increasing function of $f_i$ and $f_j$.  For larger $\beta$, $p_{ij}$ is closer to the larger of $f_i$ and $f_j$.  In the limits of $\beta \to \infty$ and $\beta \to -\infty$, $p_{ij}$ is equivalent to $\max\{f_i,f_j\}$ and $\min\{f_i,f_j\}$, respectively.  Therefore, the model in the previous section is a special case with $\alpha=1$ and $\beta\to -\infty$.

\subsection{Results for Generalized-Mean Model}

We conducted numerical simulations for various $\beta$ and $\alpha$
using Erd\H{o}s-R\'enyi random graphs with $\langle k \rangle = 150$ as
a surrogate network.  In the simulations, we controlled $f_0$ for each
$\alpha$ and $\beta$ so that the sampled network has an average degree
of $15 \pm 0.5$, i.e., about $10$ \% of the links are sampled.

\begin{figure}
\begin{center}
\includegraphics[width=\columnwidth]{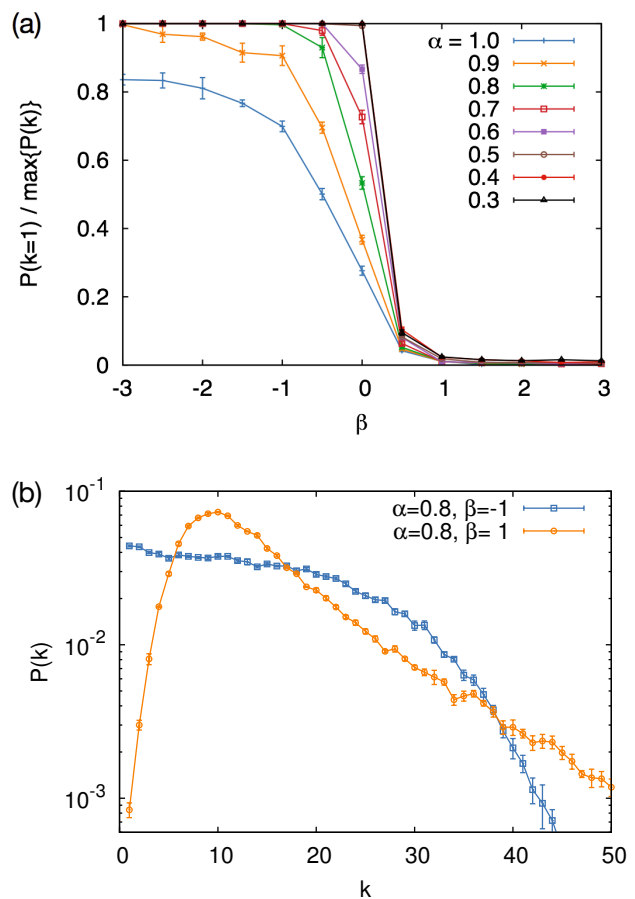}
  \caption{\label{Fig:gen_Pk}
  (Color online) (a) The fraction of nodes with degree $1$ normalized
  by the peak value of $P(k)$ in the sampled networks. The horizontal
  axis indicates the exponent of the mean $\beta$.
  (b) Typical degree distributions for positive and negative $\beta$.
  Erd\H{o}s-R\'enyi random graphs are used as the surrogate networks. The results are averaged over $5$ independent runs. }
\end{center}
\end{figure}

First we investigated whether the degree distribution is monotonically decreasing or not. For this, we define a quantity $P_1$ as $P(k=1)/\max\left\{P(k)\right\}$. If $P(k)$ is monotonically decreasing, $P_1$ must be one while it is less than one when $P(k)$ is a peaked distribution. As shown in Fig.~\ref{Fig:gen_Pk}(a), the monotonically decreasing degree distribution is realized only when $\beta \leq 0$ and for sufficiently small values of $\alpha$. The parameter range of $\alpha$ for which $P_1{=}1$ gets wider as $\beta$ decreases, indicating that for smaller values of $\beta$, nodes have low degrees more easily.

On the other hand, when $\beta>0$, there is no parameter region where monotonically decreasing degree distribution is realized as $P_1$ quickly drops to zero. This implies that in this parameter range higher degree nodes are favoured by the sampling probability function. The resulting network will consist of a number of high degree nodes. Numerical results show that $\beta=0$ is a threshold value above which monotonically decreasing $P(k)$ is precluded. Typical degree distributions for these two regions are shown in
Fig.~\ref{Fig:gen_Pk}(b).

\begin{figure}
\begin{center}
\includegraphics[width=\columnwidth]{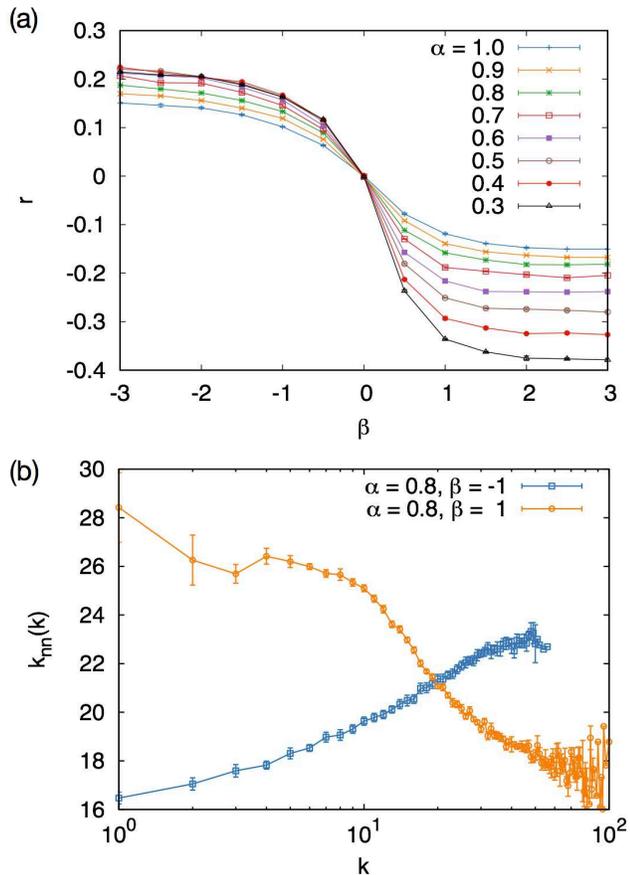}
  \caption{\label{Fig:gen_knn}
  (Color online) (a) The assortativity coefficient of the sampled
  networks as a function of $\beta$.
  (b) Typical degree correlation $k_{nn}$ for positive and negative
  $\beta$. Erd\H{o}s-R\'enyi random graphs are used as the surrogate networks. The results are averaged over $5$ independent runs. }
\end{center}
\end{figure}

Let us now investigate assortativity. The assortativity coefficient
$r$, which is the Pearson's correlation coefficient of neighboring
degrees, is plotted in Fig.~\ref{Fig:gen_knn}. The coefficient is
positive when $\beta{<}0$ while it is negative when $\beta{>}0$.
Even though affinity and original degree is independently assigned to
each node, degrees in the neighboring nodes get correlated by the
sampling. Hereafter we call this the sampling-induced assortativity.

As we have seen in the previous section, the sampling-induced assortativity originates from the correlation of $f$ of the neighboring nodes. To understand the dependency of assortativity on $\beta$, we hereafter consider the correlation of $f$. When $\beta=0$, we can analytically prove that there is no correlation of $f$ in the sampled network for any $P(f)$. The average affinity of the neighbors of a node with $f_i$, $f_{nn}(f_i)$, is calculated in a similar way as Eq.~(\ref{Eq:fnn}):
\begin{align}
  f_{nn}(f_i) &= \int_0^{1} f_j P(f_j | f_i, a_{ij}=1) df_j  \\
              &= \frac{ \int_0^1 f_j^{3/2} P(f_j) df_j }{ \int_0^1 f_j^{1/2} P(f_j) df_j },
\end{align}
which is independent of $f_i$, stating that there is no correlation of $f$ in the neighboring nodes as in the surrogate networks.

In contrast, the assortativity is modified by the sampling when $\beta \neq 0$. The differentiation of $f_{nn}$ with respect to $f_i$ is
\begin{align}\label{eq:dfnn_fi}
    \ps(f_i)^2 \frac{d}{df_i}f_{nn}(f_i) &= \int_0^1 p_{ij} P(f_j)df_j \int_0^1 \frac{dp_{ij}}{df_i} f_jP(f_j)df_j \nonumber \\
    &- \int_0^1 \frac{dp_{ij}}{df_i} P(f_j)df_j \int_0^1 p_{ij}f_jP(f_j)df_j,
\end{align}
where $\ps(f_i) = \int_0^1 p_{ij}P(f_j)df_j$. The sign of this equation can be evaluated analytically: When $\beta<0$, $df_{nn}/df_i \geq 0$ while $df_{nn}/df_i \leq 0$ when $\beta>0$. See Appendix~\ref{sect:sign_fnn} for details. Therefore, the sign of $\beta$ determines the bias of the assortativity in the sampled networks.

To conclude, we find decreasing $P(k)$ only for $\beta \leq 0$.
In the parameter region where decreasing $P(k)$ is realized, the positive correlation of $f$ is inevitable.
These results indicate that we can not conclude that the whole social network has assortative mixing even when we find assortativity in empirical network taken from one communication channel.

\section{Summary and Discussion}\label{sect:summary}

The question of general interest is to what extent ICT data can tell
us about the structure of the entire social network of people, as all
such data are incomplete and capture only a part of the whole plethora
of social relationships. While each type of service has different
features, we observe universal properties of the networks generated
from any service data. There is always a different story behind each
service but there is also commonly observed feature, namely that they
all display a decreasing degree distribution, which cannot be true for
the entire social network and hence must be attributed to the
sampling.

To investigate the effect of sampling method we have modeled how
people are using ICT communication services. The method is general
enough to be applied to various communication channels. The networks
sampled by this method robustly reproduce the stylized facts of the
ICT data: Decreasing degree distributions and assortative mixing, even
when they were absent from the original networks. Thus, the
characteristics of the sampled networks can be strongly dependent on
the sampling method, hence some properties of the original network are
hardly observed.

This result has an important implication such that properties observed
on a sampled network may not be true for the original network as
turned out to be the case with decreasing degree distribution. Hence
it can be the sampling rather than the original network that plays a
pivotal role in explaining some of the empirical network properties.
There is though a subset of users with high activity, i.e., users who
put much effort in the given ICT service. Their network properties
show features reminiscent of the true social network so that these
users can be used to some extent to reflect the properties from one
layer to the whole social network. This can also be shown in our
sampling result, where high affinity nodes show properties
characteristic for the surrogate networks, i.e., peaked degree
distribution and flat $k_{nn}$ in case of ER [see
Figs.~\ref{Fig:ER_pk_f} and~\ref{Fig:knn}(a)].

We have checked the robustness of our model by studying more general
channel selection functions and affinity distributions. Here we have
shown that there is a class of rules that result in the universally
observed single channel properties of monotonic degree distribution
and assortative mixing. This implies that the choice of the
communication channel should follow rules similar to the minimum rule,
i.e. a person may be reluctant to use a communication channel with a
friend who does not like that channel even if that is the person's
favorite. We note here that we have tested our sampling model on other
networks e.g. scale-free with similar result.
 
The sampling model presented in this paper is only one possible
mechanism for link selection. Even though obvious factors such as
three-point correlations are missing from it the qualitative agreement
between the sampling model and the empirical data is indicative.  We
did not intend to prove that the model presented here is the very
mechanism for communication channel selection but we showed that it is
enough to reproduce ICT-related observations from uncorrelated random
networks. Thus properties measured on those partial networks may not
reflect anything from the original ones, which emphasizes the
importance of the simultaneous investigation of multiple communication
channels. One of the promising ways to get a more complete picture of
human sociality is ``Reality Mining''~\cite{Eagle2006Reality,
Jo2012Spatiotemporal, Stopczynski2014Measuring}, where several
communication channels including face-to-face encounters are
simultaneously recorded. Empirical research towards such direction is
expected to reveal the relationship between the networks of different
communication channels and the way people choose among them. This will
help us to understand how much of the results from previous empirical
studies concentrating mostly on a certain communication channel can be
applied to the whole social network.

\begin{acknowledgments}
    J.~T. acknowledges financial support of AScI internship programme.
    Y.~M. appreciates hospitality at Aalto University and acknowledges
    support from CREST, JST. H.-H.~J. acknowledges financial support
    by Basic Science Research Program through the National Research 
    Foundation of Korea (NRF) grant funded by the Ministry of Education
    (2015R1D1A1A01058958). J.~K. acknowledges support from EU Grant
    No. FP7 317532 (MULTIPLEX). The systematic simulations in this
    study were assisted by OACIS~\cite{Murase2014Tool}. Partial
    support by OTKA, K112713 is gratefully acknowledged. This project
    was supported by JSPS and NRF under the Japan-Korea
    Scientific Cooperation Program. This work was supported under the
    framework of international cooperation program managed by the
    National Research Foundation of Korea (NRF-2016K2A9A2A08003695). 
    K.K. acknowledges support from Academy of Finland's COSDYN project 
    (No.  276439) and EU's Horizon 2020 FET Open RIA 662725 project IBSEN.
\end{acknowledgments}

\appendix

\section{Trial with other affinity distributions}\label{sect:other_Pf}

We have noted in the main text that the exponential affinity distribution is a special case for which the degree distribution of the sampled network starts with a constant value. This will be proven using the approximation already mentioned namely that:
\begin{align}
q(k_i|k_0,f_i)&=\binom{k_0}{k_i}\ps(f_i)^{k_i} [1-\ps(f_i)]^{k_0-k_i}\cr
              &\simeq\delta(k_0 \ps(f_i) -k_i)
\end{align}
The degree distribution is thus
\begin{align}
Q_{k_0}(k)& = \int_0^\infty q(k|k_0,f_i)P(f_i) df_i \cr
          &\simeq \int_0^\infty \delta(k_0 \ps(f_i) -k)P(f_i) df_i \cr
          &= \frac{P(f^*)}{|k_0 \ps'(f^*)|},
\end{align}
where $f^*$ is the solution of the equation $k_0 \ps(f^*) =k$.

For $P(f)= \frac{1}{f_0}e^{-f/f_0}$ the above equation gives
\begin{equation}
Q_{k_0}(k)=\begin{cases}
    \frac{1}{f_0k_0} & \mbox{if}\ k< k_0f_0\cr
    0& \mbox{if}\ k\geq k_0f_0.
\end{cases}
\end{equation}
The second case happens because $\ps(f)>f_0$.

Let us repeat the calculation for arbitrary $P(f)$ and we are
interested in the behavior of $P(k)$ for $k\ll k_0$. This latter
assumption implies $f^*\ll1$ since only nodes with low affinity have
low degree. The probability $\ps(f)$ is calculated as
\begin{align}
  \ps(f) &= \int_0^{\infty} \min\left\{f,f' \right\} P(f') df' \cr
       &= \int_0^{f} f'P(f')df' + f \int_{f}^{\infty} P(f') df'
\end{align}
The derivative is given as
\begin{equation}
  \ps'(f) = \int_f^{\infty} P(f') df'.
\end{equation}
Using this relation, the degree distribution is obtained as a function of $f^*$:
\begin{equation}
  \label{Eq:p_k0_f}
  Q_{k_0}(f^*) = \frac{ P(f^*) }{k_0 \int_{f^*}^{\infty} P(f')df'}.
\end{equation}
Since $k$ is an increasing function of $f^*$, when Eq.~(\ref{Eq:p_k0_f}) is an increasing (decreasing) function of $f^{*}$, $P(k)$ is an increasing (decreasing) function of $k$. Therefore, in order $Q_{k_0}(f^*)$ to be constant, $P(f)$ must be an exponential function.

\begin{figure}
\begin{center}
\includegraphics[width=\columnwidth]{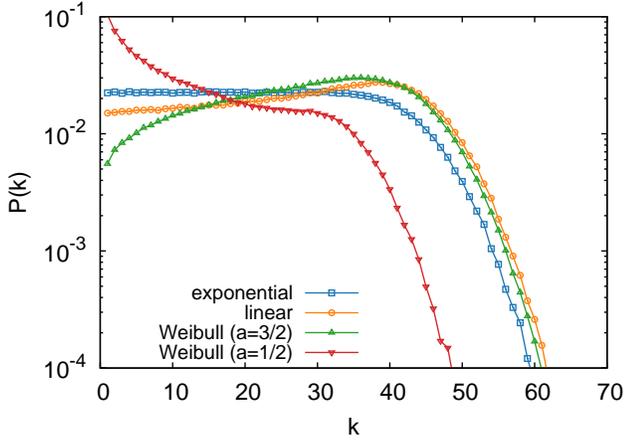}
\caption{\label{Fig:4pf_pk}
Degree distributions $P(k)$ in the sampled networks from the regular random graph of $N=10^4$ and $k_0=150$, using different functional forms of $P(f)$. The definitions of $P(f)$ are given in Table~\ref{table:different_pf}. The results are averaged over $50$ independent runs.
}
\end{center}
\end{figure}

Equation~(\ref{Eq:p_k0_f}) allows us to determine if a degree distribution function starts as decreasing or increasing. In the latter case it will definitely have a peak. As shown in Fig.~\ref{Fig:4pf_pk}, simulation results of degree distributions for different functional forms $P(f)$ are consistent with the above argument. The tested distributions are summarized in Table~\ref{table:different_pf}.

\begin{table}[htbp]
  \caption{
    Different functional forms of $P(f)$: Exponential, linear, and the Weibull distributions. Definition of the distributions are summarized as follows. Parameter $f_0$ is controlled for each distribution so that they have $\langle f \rangle = 0.3$. The equation $Q_{k_0}(f^*)$ is also calculated for $f < f_0$. We can calculate $Q_{k_0}(f^*)$ for the Weibull distributions as $\frac{af^{a-1}}{ k_0f_0^a }$, which includes an exponential distribution as a special case. $Q_{k_0}(f^*)$ is increasing (decreasing) for $a>1$ ($a<1$).
  }
  \label{table:different_pf}
  \begin{tabular}{c|ccc}
                & $P(f)$                               & $f_0$   & $Q_{k_0}(f^*)$     \\ \hline
    exponential & $\frac{1}{f_0} e^{-(f/f_0)}$         & $0.3$   & $\frac{1}{k_0f_0}$ \\
    linear      & $\frac{2}{f_0^2}(f_0-f)$ for $f<f_0$ & $0.9$   & $\frac{2}{k_0(f_0-f^*)}$  \\
    Weibull ($a=3/2$) & $\frac{a}{f_0} \left(\frac{f}{f_0}\right)^{a-1} e^{-(f/f_0)^a}$ & $0.903$ & $\frac{3\sqrt{f^*}}{2k_0f_0^{3/2} }$ \\
    Weibull ($a=1/2$) & $\frac{a}{f_0} \left(\frac{f}{f_0}\right)^{a-1} e^{-(f/f_0)^a}$ & $0.15$ &  $\frac{1}{2k_0\sqrt{f_0}\sqrt{f^*} }$
  \end{tabular}
\end{table}

\section{Sign of Equation~(\ref{eq:dfnn_fi})}\label{sect:sign_fnn}

In this appendix, we evaluate the sign of $df_{nn}/f_i$, which determines the correlation of $f$ between the neighbors in the sampled network.

Since the partial derivative of $p_{ij}$ with respect to $f_i$ is
\begin{equation}
  \frac{ \partial p_{ij} }{\partial f_i} = \frac{f_i^{\beta-1}}{
  f_i^\beta + f_j^\beta } p_{ij},
\end{equation}
Eq.~(\ref{eq:dfnn_fi}) is calculated as
\begin{align}\label{eq:dfnn_fi_2}
   & \int_0^1 p_{ij} P(f_j)df_j \int_0^1 g(f_j) f_j p_{ij} P(f_j)df_j  \\ \nonumber
  -& \int_0^1 g(f_j) p_{ij} P(f_j)df_j \int_0^1 f_j p_{ij} P(f_j)df_j                ,
\end{align}
where 
\begin{equation}
  g(f_j) \equiv \frac{f_i^{\beta-1}}{ f_i^\beta + f_j^\beta }.
\end{equation}

Here we introduce $\mu(f_j)$ as
\begin{equation}
  \mu = \int_0^{f_j} p_{ij} P(f) df.
\end{equation}
This is a positive increasing function of $f_j$. When $f_j$ changes from $0$ to $1$, $\mu$ changes from $0$ to $\int_0^1 p_{ij}P(f)df \equiv \mu_{1}$.
Using this notation, Eq.~(\ref{eq:dfnn_fi_2}) is
\begin{align}\label{eq:dfnn_fi_3}
  & \int_0^{\mu_1} d\mu(f_j) \int_0^{\mu_1} g(f_j) f_j d\mu(f_j)  \\ \nonumber
  -& \int_0^{\mu_1} g(f_j) d\mu(f_j) \int_0^{\mu_1} f_j d\mu(f_j).
\end{align}

Chebyshev integral inequality \cite{fink1984chebyshevs} states that the inequality
\begin{equation}
  \int_a^b d\mu \int_a^b fg d\mu \geq \int_a^b f d\mu \int_a^b g d\mu
\end{equation}
holds under the hypothesis that
\begin{equation}
  \left[ f(x)-f(y) \right]\left[ g(x)-g(y) \right] \geq 0
\end{equation}
for all $(x,y) \in [a,b]\times[a,b]$ and $\mu$ is a non-negative measure.
In other words, the equation holds when $f$ and $g$ have the same monotonicity.
The inverse inequality holds when $f$ and $g$ have the opposite monotonicity.
The proof is obtained by calculating the following inequality:
\begin{equation}
  1/2 \int_a^b \int_a^b [ f(x) - f(y) ][ g(x) - g(y) ] d\mu(x)d\mu(y) \geq 0.
\end{equation}

Because $g(f_j)$ is an increasing function of $f_j$ when $\beta<0$, Eq.~(\ref{eq:dfnn_fi_3}) is non-negative for an arbitrary $P(f)$.
When $\beta>0$, on the other hand, Eq.~(\ref{eq:dfnn_fi_3}) is non-positive.

\bibliography{h2jo}
\end{document}